\documentclass[twocolumn,english,aps,prd,nofootinbib]{revtex4-2}
\usepackage{lmodern}
\usepackage[T1]{fontenc}
\usepackage[latin9]{inputenc}
\usepackage{color}
\usepackage{amsmath}
\usepackage{amsthm}
\usepackage{amssymb}
\usepackage{hyperref}
 
\makeatletter

\providecommand{\tabularnewline}{\\}

\theoremstyle{plain}
\newtheorem{thm}{\protect\theoremname}
\theoremstyle{remark}
\newtheorem{rem}[thm]{\protect\remarkname}

\usepackage{enumerate}

\makeatother

\usepackage{babel}
\providecommand{\remarkname}{Remark}
\providecommand{\theoremname}{Theorem}

\begin{document}
\title{Observational tests of asymptotically flat ${\cal R}^{2}$ spacetimes
}
\author{Tao Zhu$\,$}
\email[\ \ ]{zhut05@zjut.edu.cn}

\affiliation{Institute for Theoretical Physics \& Cosmology, Zhejiang University
of Technology, Hangzhou, 310023, China ~~\\
 United Center for Gravitational Wave Physics (UCGWP), Zhejiang University
of Technology, Hangzhou, 310023, China}
\author{Hoang Ky Nguyen$\,$}
\email[\ \ ]{hoang.nguyen@ubbcluj.ro}

\affiliation{Department of Physics, Babe\c{s}--Bolyai University, Cluj-Napoca
400084, Romania}
\author{Mustapha Azreg-A\"{\i}nou$\,$}
\email[\ \ ]{azreg@baskent.edu.tr}

\affiliation{Ba\c{s}kent University, Engineering Faculty, Ba\u{g}lica Campus, 06790-Ankara,
Turkey}
\author{Mubasher Jamil$\,$}
\email[\ \ ]{mjamil@sns.nust.edu.pk}

\affiliation{School of Natural Sciences, National University of Sciences and Technology,
Islamabad 44000, Pakistan}
\date{\today}
\begin{abstract}
A novel class of Buchdahl-inspired metrics with closed-form expressions
was recently obtained based on Buchdahl's seminal work on searching
for static, spherically symmetric metrics in ${\cal R}^{2}$ gravity
in vacuo. Buchdahl-inspired spacetimes provide an interesting framework
for testing predictions of ${\cal R}^{2}$ gravity models against
observations. To test these Buchdahl-inspired spacetimes, we consider
observational constraints imposed on the deviation parameter, which
characterizes the deviation of the asymptotically flat Buchdahl-inspired
metric from the Schwarzschild spacetime. We utilize several recent
solar system experiments and observations of the S2 star in the Galactic
center and the black hole shadow. By calculating the effects of Buchdahl-inspired
spacetimes on astronomical observations both within and outside of
the solar system, including the deflection angle of light by the Sun,
gravitational time delay, perihelion advance, shadow, and geodetic
precession, we determine observational constraints on the corresponding
deviation parameters by comparing theoretical predictions with the
most recent observations. Among these constraints, we find that the
tightest one comes from the Cassini mission's measurement of gravitational
time delay.
\end{abstract}
\maketitle

\section{Introduction}

Einstein's theory of general relativity (GR) has been the most successful
theory describing the dynamics of massive objects under gravitational
effects such as motions of binary stars and planetary motions near
their host stars, predicting novel gravitational objects such as black
holes, compact stars, and gravitational waves. It remains the only
theory of gravity that passes all solar system and astronomical tests.
In the last few years, two earliest predictions of GR have been tested
and verified via black holes namely: the gravitational deflection
of light passing by a black hole resulting in the formation of black
hole shadow and the existence of gravitational waves. Recently the
shadows of black holes at the center of M87 and Milky Way galaxies
have been observed and analyzed in detail \cite{EHT1, EHT2, EHT3, EHT4}.
Astronomers have also detected numerous gravitational wave signals
generated by the merger of binary black holes of different masses
\cite{gws, gws2}. Despite these successes, some fundamental problems
remain in the foundations of GR including its renormalization and
establishing its unison with quantum mechanics thereby formulating
a set of physical laws valid for both length scales, the very small
and the very large.\vskip4pt

The accelerated cosmic expansion observed in 1998
has spurred efforts to modify GR to account for the enigmatic ``dark
energy'' component. Among the various modified theories of gravitation,
the family of $f(\mathcal{R})$ introduced by Buchdahl in early 1970s
has become an active arena of investigation in the past 25 years \cite{Buchdahl-1970,Bohmer-2023,deFelice-2010,Sotiriou-2008}.
Within this ghost-free class of theories, pure ${\cal R}^{2}$ gravity
stands out for its scale invariance nature. Attempts to incorporate
the matter sector, viz., the Glashow-Weinberg-Salam model of particle
physics into pure quadratic gravity to form a renormalizable quantum
gravity framework have been made in the form of adimensional gravity,
or ``agravity'' \cite{Einhorn-2015,Salvio-2014}.\vskip4pt

The pure ${\cal R}^{2}$ action 
\begin{equation}
S=\frac{1}{2\kappa}\int d^{4}x\sqrt{-g}\mathcal{R}^{2}+S_{M}\left(g_{\mu\nu},\psi\right) \label{eq:R2-action}
\end{equation}
can be recast using an auxiliary scalar field $\Phi$ as \cite{Sotiriou-2008}
\begin{equation}
S=\frac{1}{2\kappa}\int d^{4}x\sqrt{-g}\left[\Phi\mathcal{R}-\frac{1}{2}\Phi^{2}\right]+S_{M}\left(g_{\mu\nu},\psi\right)
\end{equation}
resulting in the field equations
\begin{align}
G_{\mu\nu} & =\frac{\kappa}{\Phi}T_{\mu\nu}-\frac{1}{4}g_{\mu\nu}\Phi+\frac{1}{\Phi}\left(\nabla_{\mu}\nabla_{\nu}\Phi-g_{\mu\nu}\square\,\Phi\right)\label{eq:G-eqn}\\
3\,\square\,\Phi & =\kappa T\label{eq:harmonic}
\end{align}
in which $\Phi$ is equal to the Ricci scalar $\mathcal{R}$. In vacuum,
viz. $T_{\mu\nu}=0$, the terms $\frac{1}{4}g_{\mu\nu}\Phi$ and $\frac{1}{\Phi}\left(\nabla_{\mu}\nabla_{\nu}\Phi-g_{\mu\nu}\square\,\Phi\right)$
act as additional ``matter'' sources to the Einstein-Hilbert field
equation, producing corrections beyond the vacuum solutions of GR.
To date, there is yet empirical evidence of a scalar degree of freedom
beside the established tensor ingredients. However, the dynamical
nature of $\Phi$, governed by the ``harmonic'' Equation \eqref{eq:harmonic},
suggests potential manifestations for $\Phi$ near mass sources in the strong
field regime. As demonstrated by one of the authors \cite{Nguyen-2022-Buchdahl},
the pure $\mathcal{R}^{2}$ field equation admits a rich host of vacuum
solutions with \emph{non-constant} Ricci scalar. The Buchdahl-inspired solutions found therein are asymptotically de Sitter (or anti-de Sitter) and entails a new (Buchdahl) parameter
$k$, accounting for the term $\frac{1}{\Phi}\left(\nabla_{\mu}\nabla_{\nu}\Phi-g_{\mu\nu}\square\,\Phi\right)$ and enabling spatial variations in $\Phi$. Notably, considerations
of a scalar degree of freedom have appeared in other contexts, such
as the scalarization in neutron stars and black holes \cite{Scalarize}.\vskip3pt

One important feature of pure $\mathcal{R}^{2}$
gravity is its limit to GR. In particular, it concerns with the existence
of a Newtonian behavior which is an essential requirement for any viable theory of
gravitation. The limit is delicate \cite{Gaume-2016}. It is because
whereas the vacuo of GR is Ricci flat, viz. ${\cal R}_{\mu\nu}=0$,
for pure ${\cal R}^{2}$ gravity the vacuum background far way from
mass sources is de Sitter (or anti-de Sitter) with a Ricci scalar
$\mathcal{R}=4\Lambda$. In \cite{Gaume-2016} it was found that,
owing to the de Sitter background, the spin-2 tensor graviton excitations
are massless instead of massive (which would have been the case if
the background were locally flat). The massless modes should effectively
carry a long-range interaction rather than a Yukawa short-range
interaction. Based on this insight, one of the present authors \cite{Nguyen-Newtonian-2023}
has recently established the emergence of a gravitational potential
with the correct Newtonian tail on a de Sitter background for the
pure $\mathcal{R}^{2}$ theory. That is to say, pure ${\cal R}^{2}$
gravity possesses a proper Newtonian limit, \emph{despite} the absence of the Einstein-Hilbert term in its action \eqref{eq:R2-action}. This finding strengthens the viability of pure ${\cal R}^{2}$ as a candidate theory of gravitation. \vskip3pt

An immediate implication is to find novel static and spherically symmetric
spacetime or black hole solutions in ${\cal R}^{2}$ theory which
was pioneered by Buchdahl \cite{Buchdahl-1962}. His work culminated
in the formulation of a second-order ordinary differential equation
for finding metric coefficients which remained unsolved for a long
time. Until recently, one of the coauthors (Nguyen) succeeded in obtaining
vacuum solutions which are asymptotically de Sitter \cite{Nguyen-2022-Buchdahl}
or asymptotically flat as a special case \cite{Nguyen-2022-Lambda0}
(with the axisymmetric extensions of the latter solution having been
proposed in \cite{2023-axisym}). Our current study shall be concerned
with this special Buchdahl-inspired solution. The metric involves
a free parameter, the Buchdahl parameter $k$, which can be interpreted
as a scalar hair, and setting $k=0$ yields a Schwarzschild spacetime/black
hole as a limiting case. In \cite{Nguyen-2022-Lambda0} it was shown
that the respective static, spherically symmetric black hole solution
has different areas of event horizon depending on the chosen value
of $k$. Specifically, the horizon area can be $0,4\pi r_{s}^{2},16\pi r_{s}^{2}$,
and divergent for the following values $k\in(-\infty,r_{s})\cup(0,\infty),k=0,k=-r_{s}$
and $k\in(-r_{s},0)$, respectively. Here $r_{s}$ is the radius of
the black hole horizon. A further investigation of this new solution
is required to understand its phenomenology. \vskip4pt

We need to clarify that the special Buchdahl-inspired
solution is also Ricci scalar flat, namely, $\Phi\rightarrow0$ everywhere
in the vacuum exterior to a mass source. Whereas the term $-\frac{1}{4}g_{\mu\nu}\Phi$
in Eq. \eqref{eq:G-eqn} is negligible in this limit, contributions from the
term $\frac{1}{\Phi}\left(\nabla_{\mu}\nabla_{\nu}\Phi-g_{\mu\nu}\square\,\Phi\right)$
persist and are encoded by the (dimensionless) Buchdahl parameter
$\tilde{k}$, defined to be $k$ normalized by $r_{s}$ \cite{Nguyen-Nontriviality-2023}.
As a higher-derivative characteristic, the value of $\tilde{k}$ is
\emph{system-dependent}. It does not have a universal value but can vary
from one system to another, depending on the composition of the matter
source. In normal conditions, such as in the Solar System, $\tilde{k}$ could be insignificant. In extreme conditions, such as around compact stars, intuitively, $\tilde{k}$ may acquire large values.\vskip4pt

In this article, we are motivated to analyze the experimental and
observational implications of the special Buchdahl-inspired metric.
We would like to see how much the new effects beyond the Schwarzschild
case affect the dynamics of particles in geodesic motion in the special
Buchdahl-inspired metric. To be more general, we consider a more general
form of Buchdahl-inspired metric by treating several new parameters
for describing the solution independently. By relying on the classical
relativistic methods, we calculate perihelion shift, gravitational
time delay, and geodesic precession of orbits, and test these results
with the solar system experiments. We also attempt to consider the
observational implications of the Buchdahl-inspired spacetime using
the observational data of S2 star orbit about the Milky Way central
black hole and to investigate the shadow of rotating solutions.\vskip4pt

Our article is structured as follows. In Sec.~\ref{sec-Buchdahl}
we provide a brief review of the general and special Buchdahl-inspired
metrics in different sets of coordinate systems. In Sec.~\ref{sec-class},
we investigate the geodesics of both massless and massive objects
in the general Buchdahl-inspired spacetime and derive in detail the
effects of the Buchdahl-inspired spacetime on observations in the
Solar System experiments, black hole shadow of M87, and the orbit
of S2 star at the Galactic center. The observational bounds on the
deviation parameter, which characterizes the deviation of the asymptotically
flat Buchdahl-inspired metric from the Schwarzschild spacetime, are
obtained by comparing the theoretical predictions with observational
data. Then, in Sec.~\ref{sec-pre}, we study a spinning object in
the general Buchdahl-inspired spacetime and derive the geodetic procession
of its spin vector, from which we obtain the constraints on the corresponding
deviation parameter by using the Gravity Probe B and lunar laser ranging
data. A brief summary of our main results and some discussions are
presented in Sec.~\ref{sec-con}.


\section{Summary of Buchdahl-inspired vacuum solutions\label{sec-Buchdahl}}

In a pioneering Nuovo Cimento work in 1962 \cite{Buchdahl-1962},
Buchdahl developed--but prematurely abandoned--a program to find
vacuum configurations for pure $\mathcal{R}^{2}$ gravity. Subsequent
advancements made by one of us, documented in Refs. \cite{Nguyen-2022-Buchdahl,Nguyen-2022-Lambda0},
completed his program and derives an exhaustive class of metric in
a compact form, to be summarized in this Section.\vskip4pt

The field equation in vacuum is 
\begin{equation}
\mathcal{R}\,\Bigl(\mathcal{R}_{\mu\nu}-\frac{1}{4}g_{\mu\nu}\mathcal{R}\Bigr)+g_{\mu\nu}\square\,\mathcal{R}-\nabla_{\mu}\nabla_{\nu}\mathcal{R}=0,
\end{equation}
which contains fourth derivatives of the metric components $g_{\mu\nu}$
in $\square\,\mathcal{R}$ and $\nabla_{\mu}\nabla_{\nu}\mathcal{R}$.
The solution in general thus involves two additional parameters. As
we shall see, they are the scalar curvature, $4\Lambda$, at spatial
infinity, and a new (Buchdahl) parameter $k$ which is of the dimension
of length. The case of $\Lambda\neq0$ is an asymptotically (anti-)de
Sitter vacuum solution, whereas the case of $\Lambda=0$ is an asymptotically
flat vacuum solution.

The body of our orbital motion study in this paper is on the asymptotically
flat vacuum solution outside of a static and spherical symmetric mass
source. Nevertheless, for completeness, we shall expose the available
representations of the metrics in this Section.

\subsection{Asymptotically de Sitter vacuum solution in standard coordinates}

In \cite{Nguyen-2022-Buchdahl}, a general Buchdahl-inspired metric
was determined to be in a compact form (with $d\Omega^{2}:=d\theta^{2}+\sin^{2}\theta\,d\phi^{2}$)
\begin{equation}
ds^{2}=e^{k\int\frac{dr}{q(r)r}}\biggl\{-\frac{p(r)q(r)}{r}dt^{2}+\frac{p(r)r}{q(r)}dr^{2}+r^{2}d\Omega^{2}\biggr\}.\label{eq:Buchdahl-metric}
\end{equation}
The variables $p$ and $q$ obey first-order ``evolution'' rules
\begin{align}
\frac{dp(r)}{dr} & =\frac{3k^{2}}{4\,r}\frac{p(r)}{q^{2}(r)},\\
\frac{dq(r)}{dr} & =\left(1-\Lambda\,r^{2}\right)\,p(r),
\end{align}
whereas the Ricci scalar is 
\begin{equation}
\mathcal{R}(r)=4\Lambda\,e^{-k\int\frac{dr}{r\,q(r)}}.
\end{equation}
The metric involves $\Lambda$, representing the scalar curvature
at spatial infinity, and $k$, the Buchdahl parameter. When $k=0$,
metric \eqref{eq:Buchdahl-metric} duly recovers the de Sitter metric
\cite{Nguyen-2022-Buchdahl}.

\subsection{Asymptotically de Sitter vacuum solution in ``canonical'' coordinates}

The metric expressed in \eqref{eq:Buchdahl-metric} contains a conformal
factor, which is inversely proportional to the Ricci scalar. In \cite{2023-axisym},
we considered making $r$ a function of $R$ such that the proper
part of the metric satisfies $g_{tt}g_{RR}=-1$. The resulting metric
is given by 
\begin{eqnarray}
ds^{2} & = & e^{k\int\frac{dR}{\Psi(R)r^{2}(R)}}\Bigl\{-\Psi(R)dt^{2}+\frac{dR^{2}}{\Psi(R)}+r^{2}(R)d\Omega^{2}\Bigr\},\nonumber \\
\end{eqnarray}
with 
\begin{eqnarray}
\Psi(R) & := & \frac{p(R)\,q(R)}{r(R)}.
\end{eqnarray}
The ``evolution'' rules now involve three functions $p(R)$, $q(R)$
and $r(R)$ 
\begin{align}
\frac{dr(R)}{dR} & =\frac{1}{p(R)},\\
\frac{dp(R)}{dR} & =\frac{3k^{2}}{4\,r(R)q^{2}(R)},\\
\frac{dq(R)}{dR} & =1-\Lambda\,r^{2}(R),
\end{align}
with the Ricci scalar being given by 
\begin{equation}
\mathcal{R}(r)=4\Lambda\,e^{-k\int\frac{dR}{\Psi(R)r^{2}(R)}}.
\end{equation}

\subsection{Asymptotically flat vacuum solution in standard coordinates}

In Ref.~\cite{Nguyen-2022-Lambda0} we further found an exact closed
analytical solution corresponding to the case of $\Lambda=0$, which
was called the special Buchdahl-inspired metric. This metric is Ricci
scalar flat, but not Ricci flat. It describes an asymptotically flat
spacetime. Hence, it is also appropriate to call it the asymptotically
flat Buchdahl-inspired metric. 

For the region $r>r_{\text{s}}$, the metric of the asymptotically
flat Buchdahl-inspired metric read 
\begin{eqnarray}
ds^{2} & = & \Bigl(1-\frac{r_{\text{s}}}{r}\Bigr)^{\tilde{k}}\biggl\{-\left(1-\frac{r_{\text{s}}}{r}\right)dt^{2}+\frac{\rho^{4}(r)\,dr^{2}}{r^{4}\bigl(1-\frac{r_{\text{s}}}{r}\bigr)}\nonumber \\
 &  & ~~~~~~~~~~~~~~~~~+\rho^{2}(r)d\Omega^{2}\biggr\},\label{eq:special-1}
\end{eqnarray}
where the function $\rho(r)$ is 
\begin{align}
\rho(r) & =\zeta\,r_{\text{s}}\frac{\left(1-\frac{r_{\text{s}}}{r}\right)^{\frac{\zeta-1}{2}}}{1-\left(1-\frac{r_{\text{s}}}{r}\right)^{\zeta}}\label{eq:special-2}
\end{align}
and the dimensionless parameters are 
\begin{equation}
\tilde{k}:=\frac{k}{r_{\text{s}}},\ \ \ \zeta:=\sqrt{1+3\tilde{k}^{2}}.\label{eq:special-3}
\end{equation}

\subsection{Asymptotically flat vacuum solution in ``q'' coordinate}

In \cite{Nguyen-2022-Lambda0}, one of us reported yet an alternative
expression for the asymptotically flat Buchdahl-inspired metric, Eq.
\eqref{eq:special-1}. For the region $q\geqslant q_{+}$ (which are
defined in Eq. (\ref{qq})): 
\begin{eqnarray}
ds^{2} & = & \left(\frac{q-q_{+}}{q-q_{-}}\right)^{\frac{\tilde{k}-1}{\zeta}}\biggl\{-\left(\frac{q-q_{+}}{q-q_{-}}\right)^{\frac{2}{\zeta}}dt^{2}+dq^{2}\nonumber \\
 &  & ~~~~~~~~~~~~~~~~+(q-q_{+})(q-q_{-})d\Omega^{2}\biggr\}
\end{eqnarray}
with 
\begin{align}
q_{\pm} & :=\frac{r_{\text{s}}}{2}\left(-1\pm\zeta\right),\ \ \ \zeta=\sqrt{1+3\tilde{k}^{2}}.\label{qq}
\end{align}

\subsection{Asymptotically flat vacuum solution in isotropic coordinate}

The metric \eqref{eq:special-1} can be transformed into an isotropic
form \cite{2023-WH}, 
\begin{eqnarray}
ds^{2} & = & \left|\frac{\bar{r}-r_{\text{s}}/4}{\bar{r}+r_{\text{s}}/4}\right|^{\frac{2}{\zeta}(\zeta+\tilde{k}-1)}\biggl\{-\left|\frac{\bar{r}-r_{\text{s}}/4}{\bar{r}+r_{\text{s}}/4}\right|^{\frac{2}{\zeta}(2-\zeta)}dt^{2}\nonumber \\
 &  & \ \ \ \ \ \ \ \ \ \ \ \ +\zeta^{2}\left(1+\frac{r_{\text{s}}}{4\bar{r}}\right)^{4}\left(d\bar{r}^{2}+\bar{r}^{2}d\Omega^{2}\right)\biggr\},
\end{eqnarray}
which is symmetric with respect to a reciprocal coordinate transformation,
per 
\begin{equation}
\frac{4\bar{r}}{r_{\text{s}}}\leftrightarrows\frac{r_{\text{s}}}{4\bar{r}}.
\end{equation}

\subsection{\label{subsec:MT-form}Asymptotically flat vacuum solution in Morris-Thorne
form}

In Ref. \cite{2023-WH}, the metric \eqref{eq:special-1}--\eqref{eq:special-3}
was brought into the Morris-Thorne form \cite{MorrisThorne-1988-1,MorrisThorne-1988-2}:
\begin{eqnarray}
ds^{2} & = & -e^{2\Phi(R)}dt^{2}+\frac{dR^{2}}{1-\frac{b(R)}{R}}+R^{2}d\Omega^{2},\label{eq:MT-1}\\
e^{2\Phi(R)} & = & y^{\frac{2}{\zeta}(\tilde{k}+1)},\label{eq:MT-2}\\
1-\frac{b(R)}{R} & = & \frac{1}{4y^{2}}\left((y^{2}+1)+\frac{\tilde{k}-1}{\zeta}(1-y^{2})\right)^{2}\geq0,\nonumber \\
~~~\label{eq:MT-3}\\
R & = & (\zeta r_{\text{s}})\frac{y^{\frac{\tilde{k}-1}{\zeta}+1}}{1-y^{2}},\label{eq:MT-4}\\
y & := & \left(1-\frac{r_{\text{s}}}{r}\right)^{\frac{\zeta}{2}}\in(0,1),\ \ \ \zeta=\sqrt{1+3\tilde{k}^{2}},\nonumber \\
\label{eq:MT-5}
\end{eqnarray}
with $\Phi(R)$ and $b(R)$ being the redshift and shape functions,
respectively. Note that the relation $y(R)$ is implicit by inverting
Eq.~\eqref{eq:MT-4}.

\subsection{A more generic Morris-Thorne form}

In Ref. \cite{2023-WEC}, we generalized the metric in \eqref{eq:MT-1}--\eqref{eq:MT-5}
by making two modifications: (i) replacing $\tilde{k}$ with $\eta$
in the redshift function (see below); (ii) treating $\tilde{k}$,
$\eta$ and $\zeta$ as \emph{independent} parameters (in contrast
to Eq. \eqref{eq:MT-5}, where $\zeta=\sqrt{1+3\tilde{k}^{2}}$.
The generalized metric is expressed as 
\begin{eqnarray}
ds^{2} & = & -e^{2\Phi(R)}dt^{2}+\frac{dR^{2}}{1-\frac{b(R)}{R}}+R^{2}d\Omega^{2}\label{eq:MT-generic-1}
\end{eqnarray}
with 
\begin{eqnarray}
e^{2\Phi(R)} & = & y^{\frac{2}{\zeta}(\eta+1)},\label{eq:MT-generic-2}\\
1-\frac{b(R)}{R} & = & \frac{1}{4y^{2}}\left((y^{2}+1)+\frac{\tilde{k}-1}{\zeta}(1-y^{2})\right)^{2}\geq0,\nonumber \\
~~\label{eq:MT-generic-3}\\
R & = & (\zeta r_{\text{s}})\frac{y^{\frac{\tilde{k}-1}{\zeta}+1}}{1-y^{2}};\ \ \ y\in(0,1).\label{eq:MT-generic-4}
\end{eqnarray}

Here we would like to present several remarks about the properties
of the solution in the generic Morris-Thorne form. 
\begin{rem}
The metric \eqref{eq:MT-generic-1}--\eqref{eq:MT-generic-4} recovers
the metric \eqref{eq:MT-1}--\eqref{eq:MT-5} when $\eta=\tilde{k}$
and $\zeta=\sqrt{1+3\tilde{k}^{2}}$. Additionally, it recovers the
Campanelli-Lousto metric in Brans-Dicke gravity \cite{Agnese-1995,Agnese-2001,Campanelli-1993,Vanzo-2012}
when $\zeta=1$. 
\end{rem}

\begin{rem}
Although the metric \eqref{eq:MT-generic-1}--\eqref{eq:MT-generic-4}
seems to have four parameters $\left\{ \tilde{k},\eta,\zeta,r_{\text{s}}\right\} $,
it effectively depends on only \emph{three} parameters: $\alpha:=\frac{\eta+1}{\zeta}$,
$\beta:=\frac{\tilde{k}-1}{\zeta}$, $r'_{\text{s}}:=\zeta r_{\text{s}}$,
per 
\begin{eqnarray}
ds^{2} & = & -e^{2\Phi(R)}dt^{2}+\frac{dR^{2}}{1-\frac{b(R)}{R}}+R^{2}d\Omega^{2},\label{eq:MT-sim-1}\\
e^{2\Phi(R)} & = & y^{2\alpha}\label{eq:MT-sim-2}\\
1-\frac{b(R)}{R} & = & \frac{1}{4y^{2}}\left((y^{2}+1)+\beta\,(1-y^{2})\right)^{2}\geq0,\label{eq:MT-sim-3}\\
R & = & r'_{\text{s}}\frac{y^{\beta+1}}{1-y^{2}}\ \ \text{for}\ \ y\in(0,1),\label{eq:MT-sim-4}
\end{eqnarray}
with the relation $y(R)$ being implicit in Eq.~\eqref{eq:MT-sim-4}. 
\end{rem}
\vskip4pt 
\begin{rem}
The Schwarzschild metric corresponds to metric \eqref{eq:MT-sim-1}--\eqref{eq:MT-sim-4}
when $\alpha=1$, $\beta=-1$ and $\zeta=1$. Meanwhile, the asymptotically
flat Buchdahl-inspired metric corresponds to $\left\{ \alpha=\frac{\tilde{k}+1}{\sqrt{1+3\tilde{k}^{2}}},\beta=\frac{\tilde{k}-1}{\sqrt{1+3\tilde{k}^{2}}}\right\} $,
obeying the relation $\alpha^{2}+\alpha\beta+\beta^{2}=1$. 
\end{rem}
\vskip4pt 
\begin{rem}
\label{rem:WH-formation}Regardless of $\alpha,$ when $\beta<-1$,
metric \eqref{eq:MT-sim-1}--\eqref{eq:MT-sim-4} yields a wormhole
because the function $R(y)$ in \eqref{eq:MT-sim-4} produces a minimum
at $y_{0}=\sqrt{\frac{\beta+1}{\beta-1}}\in(0,1)$. The two (symmetric)
asymptotically flat sheets that get glued together at the ``throat''
$y_{0}$ are both defined in the range $y\in[y_{0},1)$. 
\end{rem}

\section{geodesics and classical tests of the general Buchdahl-inspired metrics\label{sec-class}}

In this section, we will present the geodesics evolution of a massive/massless
particle orbiting the Buchdahl-inspired metrics. From the geodesic
evolution, we are able to calculate some observational quantities
so that we can use data to constrain them. We claim that the data
from the solar system, experiments can make stronger constraints than
those obtained upon using the stellar stars orbiting the supermassive
black hole in the galactic center, so we will consider the solar system
experiments first.

\subsection{Geodesics in the general Buchdahl-inspired metrics}

In this subsection, we give the general Buchdahl-inspired metrics
in the standard coordinates and consider the geodesics of both massless
and massive objects in this metric. Different from the metric given
in (\ref{eq:special-1}), here we treat the parameters $\tilde{k}$,
$\eta$, and $\zeta$ as independent parameters. Using the coordinates
$(t,R,\theta,\phi)$, the metric is given by 
\begin{eqnarray}
ds^{2} & = & -e^{2\Phi(R)}dt^{2}+\frac{dR^{2}}{1-\frac{b(R)}{R}}+R^{2}d\Omega^{2}.\label{R_coors}
\end{eqnarray}
In the weak field approximation, this metric can be expressed as 
\begin{eqnarray}
ds^{2} & \simeq & -\left(1-\frac{(1+\eta)r_{s}}{R}+\frac{(1+\eta)(\tilde{k}+\eta)r_{s}^{2}}{2R^{2}}\right)dt^{2}\nonumber \\
 &  & +\left(1+\frac{(1-\tilde{k})r_{s}}{R}\right)dR^{2}+R^{2}d\Omega^{2}.\label{weak}
\end{eqnarray}
The radius $r_{s}$ can be related to the ADM mass of the solution
via $r_{s}=2GM$ with $M$ being the ADM mass and $G$ being the gravitational
constant. Comparing this weak field expansion with the Newtonian limit,
one can relate $G$ with the Newtonian gravitational constant $G_{{\rm N}}$
as $(1+\eta)G=G_{{\rm N}}$. 

Let us first consider the evolution of a particle in the general Buchdahl-inspired
metric (\ref{R_coors}). For a massive/massless particle, if one ignores
self-gravitational effects, its evolution is governed by the following
geodesics 
\begin{equation}
\frac{d^{2}x^{\mu}}{d\lambda^{2}}+\Gamma_{\nu\rho}^{\mu}\frac{dx^{\nu}}{d\lambda}\frac{dx^{\rho}}{d\lambda}=0,
\end{equation}
where $\lambda$ denotes the affine parameter of the geodesics and
$\Gamma_{\nu\rho}^{\mu}$ represent the Christoffel symbols of the
general Buchdahl-inspired metric. Considering the general Buchdahl-inspired
metric is static and spherically symmetric, it has two Killing vectors,
$\xi_{t}^{\mu}=\{\partial_{t},0,0,0\}$ and $\xi_{\phi}^{\mu}=\{0,0,\partial_{\phi},0\}$,
which leads to two constants of motion $E$ and $L$ (conserved energy
and angular momentum), i.e., 
\begin{eqnarray}
E=-g_{\mu\nu}\xi_{t}^{\mu}\frac{dx^{\nu}}{d\lambda}=-g_{tt}\frac{dt}{d\lambda},\label{energy}\\
L=g_{\mu\nu}\xi_{\phi}^{\mu}\frac{dx^{\nu}}{d\lambda}=g_{\phi\phi}\frac{d\phi}{d\lambda}.\label{angularM}
\end{eqnarray}
For geodesics, we also have $g_{\mu\nu}\frac{dx^{\mu}}{d\lambda}\frac{dx^{\nu}}{d\lambda}=\varepsilon$
with $\varepsilon=-1$ for timelike geodesics which describes evolution
of massive particle and $\varepsilon=0$ for null geodesics which
describes the evolution of massless particle. Then using (\ref{energy})
and (\ref{angularM}), one obtains 
\begin{eqnarray}
 &  & g_{RR}\left(\frac{dR}{d\lambda}\right)^{2}+g_{\theta\theta}\left(\frac{d\theta}{d\lambda}\right)^{2}\nonumber \\
 &  & ~~~~~~~~~~=\varepsilon-g_{tt}\left(\frac{dt}{d\lambda}\right)^{2}-g_{\phi\phi}\left(\frac{d\phi}{d\lambda}\right)^{2}\nonumber \\
 &  & ~~~~~~~~~~=\varepsilon-\frac{E^{2}}{g_{tt}}-\frac{L^{2}}{g_{\phi\phi}}.
\end{eqnarray}
Without generality, we consider the evolution of the particle in the
equatorial plane, i.e., we can set $\theta=\pi/2$ and $d\theta/d\lambda=0$.
Then one can simplify the above equation into the form 
\begin{eqnarray}
\left(\frac{dR}{d\lambda}\right)^{2}=E^{2}-V_{{\rm eff}}(R),\label{rdot}
\end{eqnarray}
where $V_{{\rm eff}}(R)$ denotes the effective potential of the particle,
\begin{eqnarray}
V_{{\rm eff}}(R)=E^{2}-\left(\varepsilon-\frac{E^{2}}{g_{tt}}-\frac{L^{2}}{g_{\phi\phi}}\right)\frac{1}{g_{RR}}.
\end{eqnarray}
For later convenience, it is useful to give the derivative of $r$
with respect to $\phi$, which can be obtained by using (\ref{angularM})
in Eq.~(\ref{rdot}) and is given by 
\begin{eqnarray}
\left(\frac{dR}{d\phi}\right)^{2}=\left(\varepsilon-\frac{E^{2}}{g_{tt}}-\frac{L^{2}}{g_{\phi\phi}}\right)\frac{g_{\phi\phi}^{2}}{L^{2}g_{RR}}.\label{drdphi}
\end{eqnarray}
This equation is the starting point for later calculations of the
light deflection angle, gravitational time delay, and perihelion advance
in the general Buchdahl-inspired metric in the following subsections.

\subsection{Light deflection angle}

The precise measurements of the deflection of the light passing by
the Sun play an essential role in the establishment of GR. These data
can also be used for constraining any possible derivation of the deflection
angle in many modified gravities from that in GR. Here our purpose
is to calculate the possible effects of the general Buchdahl-inspired
metric on the deflection angle of the light and then constrain them
by using the most recent measurements.

For the propagation of the light in the general Buchdahl-inspired
metric (\ref{R_coors}), we have $\varepsilon=0$. Introducing the
impact parameter 
\begin{eqnarray}
b\equiv\frac{L}{E},
\end{eqnarray}
Eq.~(\ref{drdphi}) can be transformed into 
\begin{eqnarray}
\frac{d\phi}{dR}=\pm\sqrt{\frac{g_{RR}}{g_{\phi\phi}}}\left(-\frac{g_{\phi\phi}}{b^{2}g_{tt}}-1\right)^{-1/2},\label{dphidr}
\end{eqnarray}
where $\pm$ represents the cases with increasing and decreasing $R$,
respectively. In general, for a bending light that does not fall into
the object described by the general Buchdahl-inspired metric (\ref{R_coors}),
the range of the allowed $R$ is determined by the condition $\frac{dR}{d\lambda}\geq0$.
In the general Buchdahl-inspired metric, this implies the allowed
range of $R$ should be $R_{0}\leq R<+\infty$ with $R_{0}$ denoting
the closest approach of the light to the Sun. $R_{0}$ is a root of
$\frac{dR}{d\lambda}=0$, thus one has 
\begin{eqnarray}
b^{2}=-\frac{g_{\phi\phi}(R_{0})}{g_{tt}(R_{0})}.\label{r0}
\end{eqnarray}
Then the deflection of the angle of the light can be calculated by
using 
\begin{eqnarray}
\Delta\phi=2\int_{R_{0}}^{+\infty}\frac{d\phi}{dR}dR-\pi.
\end{eqnarray}
Since we are considering the deflection of the light by the Sun, it
is convenient to employ the weak field approximation by expanding
the above integral in terms of $r_{s}/R$, which gives 
\begin{eqnarray}
\Delta\phi\simeq\frac{4GM}{R_{0}}\frac{2+\eta-k}{2}\simeq\frac{4G_{{\rm N}}M}{R_{0}}\frac{2+\eta-k}{2(1+\eta)},
\end{eqnarray}
where $G_{{\rm N}}=(1+\eta)G$ is the Newtonian gravitational constant.
For a special Buchdahl-inspired metric, as given in (\ref{eq:special-1})
with $\eta=\tilde{k}$ and $\zeta=\sqrt{1+3\tilde{k}^{2}}$, one has
\begin{eqnarray}
\Delta\phi\simeq\frac{4G_{{\rm N}}M}{R_{0}}(1-\tilde{k}),
\end{eqnarray}
where $G_{{\rm N}}=(1+\tilde{k})G$ is the Newtonian gravitational
constant for the Buchdahl-inspired metric.

Now we consider the light deflected by the Sun. One can express the
deflection angle $\Delta\phi$ in terms of the $\Delta\phi^{{\rm GR}}=1.75''$
as 
\begin{eqnarray}
\frac{\Delta\phi}{\Delta\phi^{{\rm GR}}}=1-\frac{\eta+\tilde{k}}{2(1+\eta)}.\label{47}
\end{eqnarray}
The deflection angles of the lights from distance sources by the Sun
have been measured in many experiments in the past 100 years. Currently,
the most precise measurement was carried out by using the technique
of the very-long baseline interferometry \cite{Fomalont:2009zg}.
By using the result of this measurement, one can constrain the parameter
$\frac{\eta+\tilde{k}}{2(1+\eta)}$ in the general Buchdahl-inspired
metric to be 
\begin{eqnarray}
-5.0\times10^{-5}<\frac{\eta+\tilde{k}}{2(1+\eta)}<2.5\times10^{-4}\;\;(68\%\;{\rm C.L.}).
\end{eqnarray}
For a special Buchdahl-inspired metric (\ref{eq:special-1}) with
$\eta=\tilde{k}$ and $\zeta=\sqrt{1+3\tilde{k}^{2}}$, this constraint
leads to a constraint on the parameter $\tilde{k}$ as 
\begin{eqnarray}
-5.0\times10^{-5}<\tilde{k}<2.5\times10^{-4}\;\;(68\%\;{\rm C.L.}).
\end{eqnarray}

\subsection{ Gravitational Time Delay}

Gravitational time delay is an important phenomenon in that lights
or radio waves can take more time to travel if they pass by a massive
object, like the Sun or a planet. This phenomenon can be precisely
measured by sending a radar signal from Earth or a spacecraft passing
through the Sun and reflecting off another planet or spacecraft. The
effects of the general Buchdahl-inspired metric (\ref{R_coors}) on
the gravitational time delay can be derived from Eq.~(\ref{dphidr}),
from which one obtains 
\begin{eqnarray}
\frac{dt}{dR} & = & \frac{dt}{d\phi}\frac{d\phi}{dR}=\frac{d\phi}{dR}\frac{dt/d\lambda}{d\phi/d\lambda}\nonumber \\
 & = & \pm\frac{1}{b}\sqrt{-\frac{g_{RR}}{g_{tt}}}\left(\frac{1}{b^{2}}+\frac{g_{tt}}{g_{\phi\phi}}\right)^{-1/2}.
\end{eqnarray}
Considering a radio wave travels from the Sun to the point $R_{A}$,
the time spent during this process can be calculated from the integral
\begin{eqnarray}
t(R_{A})=\frac{1}{b}\int_{R_{0}}^{R_{A}}\sqrt{-\frac{g_{RR}}{g_{tt}}}\left(\frac{1}{b^{2}}+\frac{g_{tt}}{g_{\phi\phi}}\right)^{-1/2}dR.
\end{eqnarray}
Here $R_{0}$ is the closest approach of the radio wave to the Sun
which can be determined by Eq.~(\ref{r0}). In the weak field approximation,
one has 
\begin{eqnarray}
t(r_{A}) & \simeq & \sqrt{R_{A}^{2}-R_{0}^{2}}+G_{{\rm N}}M\sqrt{\frac{R_{A}-R_{0}}{R_{A}+R_{0}}}\nonumber \\
 &  & +\frac{2+\eta-\tilde{k}}{1+\eta}G_{{\rm N}}M{\rm arccosh}\left(\frac{R_{A}}{R_{0}}\right).
\end{eqnarray}
The first term is the travel time of light in flat spacetime and the
rest part contains both the contributions to the travel time in Schwarzschild
metric and the new effects of the general Buchdahl-inspired metric.
When $\tilde{k}=0=\eta$, the above expression exactly reduces to
the Schwarzschild result.

There are two different cases in the experiments of measuring the
gravitational time delay by sending a radar wave from Earth or spacecraft
and then reflecting off another planet or spacecraft. One is the inferior
conjunction case, in which the planet (or spacecraft, denoted by B),
which reflects the radar signal, is located between the Earth (or
spacecraft, denoted by A) and the Sun. The calculation of the time
delay due to the general Buchdahl-inspired metric of this case is
simple and can be obtained by 
\begin{eqnarray}
\Delta t_{I}\simeq2\frac{2+\eta-\tilde{k}}{1+\eta}G_{{\rm N}}M\ln\frac{R_{A}}{R_{B}}=\Delta t_{I}^{{\rm GR}}\frac{2+\eta-\tilde{k}}{2(1+\eta)},
\end{eqnarray}
alternatively, 
\begin{equation}
\frac{\Delta t_{I}}{\Delta t_{I}^{{\rm GR}}}=\frac{2+\eta-\tilde{k}}{2(1+\eta)}.\label{54}
\end{equation}

Another one is the superior conjunction case, in which the planet
that reflects the radar signal and the Earth are on opposite sides
of the Sun. For this case, the gravitational time delay is given by
\begin{eqnarray}
\Delta t_{S} & \simeq & \frac{2+\eta-\tilde{k}}{2(1+\eta)}4G_{{\rm N}}M\left[1+\ln\frac{4R_{A}R_{B}}{R_{0}^{2}}\right].\label{ts}
\end{eqnarray}

The most precise results related to the gravitational time delay were
obtained from the Cassini experiments \cite{cassini}. This result
ruled out many modified gravity theories that predicted larger deviations
from GR. Here we would like to use its results to constrain the parameter
$\tilde{k}$ and $\eta$ in the general Buchdahl-inspired metric.
The Cassini experiment was conducted in June 2002, and the test of
the gravitational time delay was achieved in the measurement of the
frequency shift of radio waves to and from the Cassini spacecraft
as they passed near the Sun. In the superior conjunction case, the
relative change in the frequency is related to the time delay $\Delta t_{S}$
via 
\begin{eqnarray}
\delta\nu=\frac{\nu(t)-\nu_{0}}{\nu_{0}}=\frac{d}{dt}\Delta t_{S},
\end{eqnarray}
where $\nu_{0}$ denotes the frequency of the radio waves emitted
from the Earth and $\nu(t)$ is the frequency of the radio wave reflected
back to the Earth at $t$. Using Eq.~(\ref{ts}), one has 
\begin{eqnarray}
\delta\nu\simeq-\frac{2+\eta-\tilde{k}}{2(1+\eta)}\frac{8G_{{\rm N}}M}{R_{0}}\frac{dR_{0}(t)}{dt}=\frac{2+\eta-\tilde{k}}{2(1+\eta)}\delta\nu^{{\rm GR}}.
\end{eqnarray}
Alternatively, 
\begin{equation}
\frac{\delta\nu}{\delta\nu^{{\rm GR}}}=\frac{2+\eta-\tilde{k}}{2(1+\eta)}.
\end{equation}

Using the measurement performed in the Cassini experiment \cite{cassini},
one can constrain the parameters $\tilde{k}$ and $\eta$ in the general
Buchdahl-inspired metric to be 
\begin{eqnarray}
-4.4\times10^{-5}<\frac{\eta+\tilde{k}}{2(1+\eta)}<2\times10^{-6}.
\end{eqnarray}
This constraint is stronger than that obtained by the observations
of the deflection angle.

For a special Buchdahl-inspired metric (\ref{eq:special-1}) with
$\eta=\tilde{k}$ and $\zeta=\sqrt{1+3\tilde{k}^{2}}$, the above
constraint leads to a constraint on the parameter $\tilde{k}$ as
\begin{eqnarray}
-4.4\times10^{-5}<\tilde{k}<2\times10^{-6}.
\end{eqnarray}

\subsection{Perihelion Advance}

Now we consider the orbit's perihelion advance for a massive particle
moving in the general Buchdahl-inspired metric. For a massive particle,
we have $\varepsilon=-1$. We still start with Eq.~(\ref{drdphi})
by introducing a new variable $x=1/R$, which leads to 
\begin{eqnarray}
\left(\frac{dx}{d\phi}\right)^{2}=x^{4}\left[-1-\frac{E^{2}}{g_{tt}}-\frac{L^{2}}{g_{\phi\phi}}\right]\frac{g_{\phi\phi}^{2}}{L^{2}g_{RR}}.
\end{eqnarray}
Taking the derivative on both sides of the above equation with respect
to $\phi$ and expanding the equation about the small parameter $r_{s}$
(in the weak field approximation), one obtains 
\begin{eqnarray}
\frac{d^{2}x}{d\phi^{2}}+x-\frac{G_{{\rm N}}M}{L^{2}} & \simeq & 3G_{{\rm N}}Mx^{2}-3\frac{\eta+\tilde{k}}{1+\eta}G_{{\rm N}}Mx^{2}\nonumber \\
 &  & +2\frac{\tilde{k}+\eta}{1+\eta}\frac{G_{{\rm N}}^{2}M^{2}}{L^{2}}x.
\end{eqnarray}
The right-hand side of the above equation can be treated as perturbations
to Newtonian gravity, which contains two parts. The first part $3G_{{\rm N}}Mx^{2}$
represents the correction from the Schwarzschild metric in GR and
the last two terms (the second and the third term on the right-hand
side of the above equation) are the new effects from the general Buchdahl-inspired
metric. When the perturbations are absent, the above equation has
an exact solution $x_{0}(\phi)$ for a bounded orbit, 
\begin{eqnarray}
x_{0}(\phi)=\frac{G_{{\rm N}}M}{L^{2}}(1+e\cos\phi),\label{elliptical}
\end{eqnarray}
which describes an elliptical orbit with the eccentricity $e$. The
perturbations from GR and the general Buchdahl-inspired metric lead
to the small derivation to the exact elliptical orbit. Thus we can
write the orbit which contains the effects of the perturbations in
the form of 
\begin{eqnarray}
x(\phi)=x_{0}(\phi)+x_{1}(\phi),
\end{eqnarray}
where $x_{0}$ is the elliptical orbit with the eccentricity $e$
given by Eq.~(\ref{elliptical}) and $x_{1}(\phi)$ is the small
corrections to the elliptical orbit, which satisfies 
\begin{eqnarray}
\frac{d^{2}x_{1}}{d\phi^{2}}+x_{1} & \simeq & 3G_{{\rm N}}Mx_{0}^{2}-3\frac{\eta+\tilde{k}}{1+\eta}G_{{\rm N}}Mx_{0}^{2}\nonumber \\
 &  & +2\frac{\tilde{k}+\eta}{1+\eta}\frac{G_{{\rm N}}^{2}M^{2}}{L^{2}}x_{0}.
\end{eqnarray}
Using $x_{0}(\phi)$ in Eq.~(\ref{elliptical}), one gets 
\begin{eqnarray}
\frac{d^{2}x_{1}}{d\phi^{2}}+x_{1}\simeq A_{0}+A_{1}\cos\phi+A_{2}\cos^{2}\phi,
\end{eqnarray}
where 
\begin{eqnarray}
A_{0} & = & \frac{3+2\eta-\tilde{k}}{1+\eta}\frac{G_{{\rm N}}^{3}M^{3}}{L^{4}},\\
A_{1} & = & \frac{6+2\eta-4\tilde{k}}{1+\eta}e\frac{G_{{\rm N}}^{3}M^{3}}{L^{4}},\\
A_{2} & = & 3\frac{1-\tilde{k}}{1+\eta}e^{2}\frac{G_{{\rm N}}^{3}M^{3}}{L^{4}}.
\end{eqnarray}
The solution of $x_{1}$ is given by 
\begin{eqnarray}
x_{1}=A_{0}+\frac{A_{2}}{2}-\frac{A_{2}}{6}\cos(2\phi)+\frac{A_{1}}{2}\phi\sin\phi.
\end{eqnarray}
Only the last term contributes to the perihelion advance of a massive
particle moving in the general Buchdahl-inspired metric. For this
reason, we can drop other terms and write the solution of $x(\phi)$
as 
\begin{eqnarray}
x & \simeq & \frac{G_{{\rm N}}M}{L^{2}}(1+e\cos\phi)+\frac{A_{1}}{2}\phi\sin\phi\nonumber \\
 & = & \frac{G_{{\rm N}}M}{L^{2}}\left[1+e\cos\left(\phi-\frac{\delta\phi_{0}}{2\pi}\phi\right)\right],
\end{eqnarray}
where 
\begin{eqnarray}
\delta\phi_{0}\simeq\frac{6\pi G_{{\rm N}}^{2}M^{2}}{L^{2}}\left(1-\frac{2}{3}\frac{\eta+\tilde{k}}{1+\eta}\right).
\end{eqnarray}
This expression represents the angular shift of the perihelia per
orbit.

For an ellipse described by Eq.~(\ref{elliptical}), one can relate
the angular momentum $L$ of the massive particle to the semi-major
axis $a_{0}$ of the ellipse as 
\begin{eqnarray}
a_{0}=\frac{L^{2}}{G_{{\rm N}}M(1-e^{2})}.
\end{eqnarray}
Thus one obtains 
\begin{eqnarray}
\Delta\phi=\frac{6\pi G_{{\rm N}}M}{a_{0}(1-e^{2})}\left(1-\frac{2}{3}\frac{\eta+\tilde{k}}{1+\eta}\right).
\end{eqnarray}
Alternatively, 
\begin{equation}
\frac{\Delta\phi}{\Delta\phi^{{\rm GR}}}\simeq1-\frac{2}{3}\frac{\eta+\tilde{k}}{1+\eta}.\label{75}
\end{equation}
It is evident that the above expression reduces to the Schwarzschild
result by taking $\tilde{k}=0=\eta$.

There are several observations of the perihelion advance that can
be used to constrain the parameters $\eta$ and $\tilde{k}$ in the
general Buchdahl-inspired metric. Now we consider three different
observations related to the phenomenon of the perihelion advance in
very different scales, i.e., the perihelion advances of the laser-ranged
satellites orbiting the earth \cite{Lucchesi:2010zzb}, of Mercury
orbiting the Sun \cite{Mercury}, and of the S2 star orbiting the
supermassive black hole in the central region of our Milky Way galaxy
\cite{s2}.

Let us first consider the measured perihelion advance of LAGEOS satellites
around the Earth. Using 13 years of tracking data of the LAGEOS satellites,
the precession of the periapsis of the LAGEOS II satellite was measured
to be \cite{Lucchesi:2010zzb} 
\begin{eqnarray}
\frac{\Delta\phi}{\Delta\phi^{{\rm GR}}}=1+(0.28\pm2.14)\times10^{-3}.
\end{eqnarray}
From this result, one obtains 
\begin{eqnarray}
-1.8\times10^{-3}<\frac{1}{2}\frac{\eta+\tilde{k}}{1+\eta}<1.4\times10^{-3}.
\end{eqnarray}
This bound corresponds to a constraint on the parameter $\tilde{k}$
\begin{eqnarray}
-1.8\times10^{-3}<\tilde{k}<1.4\times10^{-3}
\end{eqnarray}
for the special Buchdahl-inspired metric with $\eta=\tilde{k}$ and
$\zeta=\sqrt{1+3\tilde{k}^{2}}$.

We now turn to consider the observation of the anomalous perihelion
advance for Mercury. The most accurate measurement of the perihelion
advance was performed by the MESSENGER mission \cite{Mercury}, which
measures the perihelion advance for Mercury to be 
\begin{eqnarray}
\Delta\phi=(42.9799\pm0.009)''/{\rm century}.
\end{eqnarray}
With this measurement, the bound on the $\frac{\eta+\tilde{k}}{1+\eta}$
arising from the general Buchdahl metric can be computed using the
experimental error $0.009''/{\rm century}$, which yields 
\begin{eqnarray}
-1.6\times10^{-5}<\frac{1}{2}\frac{\eta+\tilde{k}}{1+\eta}<1.6\times10^{-5}.
\end{eqnarray}
This bound is better than that from the LAGEOS satellites by two orders
of magnitude. Again, from the above constraint, one can get the constraint
on the parameter $\tilde{k}$ for the special Buchdahl-inspired metric
with $\eta=\tilde{k}$ and $\zeta=\sqrt{1+3\tilde{k}^{2}}$, i.e.,
\begin{eqnarray}
-1.6\times10^{-5}<\tilde{k}<1.6\times10^{-5}.
\end{eqnarray}
Here we would like to mention that the measurement of the anomalous
perihelion advance for Mercury will be improved significantly in the
near future from the joint European-Japanese BepiColombo project,
which was launched in October 2018 \cite{Will:2018mcj, Bepi}. It
is expected that this mission will improve the accuracy of the perihelion
advance to be $10^{-4}$ as/century. This accuracy is one order of
magnitudes better than the current accuracy of MESSENGER mission \cite{Mercury}.
Thus, with the BepiColombo project, we expect to improve the constraints
on the parameters $\frac{\eta+\tilde{k}}{1+\eta}$ arising from the
general Buchdahl-inspired metric or the parameter $\tilde{k}$ in
the special Buchdahl-inspired spacetime to $-10^{-6}\lesssim\frac{\eta+\tilde{k}}{1+\eta}\;{\rm or}\;\tilde{k}\lesssim10^{-6}$,
which is much more restricted than that obtained from the Cassini
experiment.

At last, let us consider the S2 star orbiting the central black hole
of the Milky Way galaxy. Comparing the above two measurements, the
observations of the S2 star provide a very different environment to
test gravity in the strong gravity regime. The Schwarzschild precession
of the S2 star was measured recently by the GRAVITY collaboration
\cite{s2}, which gives 
\begin{eqnarray}
\frac{\Delta\phi}{\Delta\phi^{{\rm GR}}}=1.1\pm0.19,\label{82}
\end{eqnarray}
where 
\begin{eqnarray}
\Delta\phi^{{\rm GR}}=12'.
\end{eqnarray}
per orbit period from the prediction of GR. For the effect of the
general Buchdahl-inspired metric on the precession, this observation
leads to 
\begin{eqnarray}
-0.21<\frac{1}{2}\frac{\eta+\tilde{k}}{1+\eta}<0.067,
\end{eqnarray}
which corresponds to 
\begin{eqnarray}
-0.21<\tilde{k}<0.067\label{S2-star}
\end{eqnarray}
for the special Buchdahl-inspired metric with $\eta=\tilde{k}$ and
$\zeta=\sqrt{1+3\tilde{k}^{2}}$.

\subsection{Including Rotation: Shadow Investigation}

An exact, up to a conformal factor, stationary axisymmetric vacuum
solution for pure $R^{2}$ gravity was derived in~\cite{2023-axisym} 
\begin{widetext}
\begin{equation}
ds^{2}=A(q,\theta;a)\Big[-\frac{\Delta(q)-a^{2}\sin^{2}\theta}{\rho^{2}}dt^{2}+\frac{\rho^{2}}{\Delta(q)}dq^{2}+\rho^{2}d\theta^{2}+\frac{2a\sin^{2}\theta}{\rho^{2}}[\Delta(q)-r^{2}(q)-a^{2}]dt\,d\phi+\frac{\Sigma}{\rho^{2}}\sin^{2}\theta d\phi^{2}\Big],\label{rotation}
\end{equation}
\end{widetext}

where $r^{2}(q)=(q-q_{+})^{\frac{2q_{+}}{q_{+}-q_{-}}}(q-q_{-})^{\frac{-2q_{-}}{q_{+}-q_{-}}}$,
$\rho^{2}(q,\theta)=r^{2}(q)+a^{2}\cos^{2}\theta$, $\Delta(q)=(q-q_{+})(q-q_{-})+a^{2}$,
and $\Sigma(q,\theta)=[r^{2}(q)+a^{2}]^{2}-\Delta(q)a^{2}\sin^{2}\theta$.
The conformal factor $A(q,\theta;a)$, not needed for shadow investigation,
was determined numerically. The remaining parameters are given by
\begin{equation}
q_{+}=\frac{r_{s}}{2}[\sqrt{1+3\tilde{k}^{2}}-1],\quad q_{-}=-\frac{r_{s}}{2}[\sqrt{1+3\tilde{k}^{2}}+1].\label{rotation2}
\end{equation}

Using the Event Horizon Telescope Collaboration results~\cite{EHT1,EHT2,EHT3,EHT4},
we modeled the central black hole M87{*} by the rotating metric~\eqref{rotation},
depending on the mass $M=(1+\tilde{k})r_{s}/2$, rotation parameter
$a$, and the dimensionless parameter $\tilde{k}$. Considering the
shadow angular size and assuming $M$ and $a$ parameters are those
of M87{*}, we obtained 
\begin{equation}
-0.155\leq\tilde{k}\leq0.004\,.\label{rotation3}
\end{equation}

\section{Geodetic precession of spinning objects in the general Buchdahl-inspired
metric\label{sec-pre}}

In this section, we calculate the geodetic precession of spinning
objects in the general Buchdahl-inspired metric. In curved spacetime,
the evolution of a spinning particle follows two equations, the geodesics
equation 
\begin{eqnarray}
\frac{du^{\mu}}{d\lambda}+\Gamma_{\nu\lambda}^{\mu}u^{\nu}u^{\lambda}=0,
\end{eqnarray}
and the parallel transport equation, 
\begin{eqnarray}
\frac{ds^{\mu}}{d\lambda}+\Gamma_{\nu\lambda}^{\mu}s^{\nu}u^{\lambda}=0,\label{spin_eq}
\end{eqnarray}
where $u^{\mu}=dx^{\mu}/d\lambda$ is the four-velocity of the particle
and $s^{\mu}$ denotes the four-spin vector. $u^{\mu}$ and $s^{\nu}$
satisfy the following orthogonal condition 
\begin{eqnarray}
u^{\mu}s_{\mu}=0.
\end{eqnarray}
The four-spin vector also satisfies the normalization condition 
\begin{eqnarray}
s^{\mu}s_{\mu}=1.
\end{eqnarray}

Without generality, we consider the evolution of the particle in the
equatorial plane, i.e., we can set $\theta=\pi/2$ and $d\theta/d\lambda=0$.
Let us study the test spinning particle moves in a circular orbit,
i.e., $\dot{R}=0$, and its four-velocity $u^{\mu}$ can be expressed
as 
\begin{eqnarray}
u^{t}=\dot{t}=-\frac{E}{g_{tt}},\;\;\;\;u^{\phi}=\dot{\phi}=\frac{l}{g_{\phi\phi}}.
\end{eqnarray}
Then the angular velocity of the spinning particle is written as 
\begin{eqnarray}
\Omega=\frac{u^{\phi}}{u^{t}}=-\frac{l}{E}\frac{g_{tt}}{g_{\phi\phi}}.
\end{eqnarray}

\begin{table*}[!t]
\caption{Summary of estimates for bounds of the parameter $\frac{\eta+\tilde{k}}{1+\eta}$
arising in the general Buchdahl-inspired metric (\ref{R_coors}) and
the parameter $\tilde{k}$ in the special Buchdahl-inspired metric
from several observations.}
\label{table}
\begin{ruledtabular}
\begin{tabular}{ccc}
Experiments/ Observations & Constraints on $\frac{\eta+\tilde{k}}{2(1+\eta)}$ and $\tilde{k}$ & Datasets\tabularnewline
\hline 
Light deflection & $(-5.0,\;25)\times10^{-5}$ & VLBI observation of quasars \cite{Fomalont:2009zg}\tabularnewline
Time delay & $(-44,\;2)\times10^{-6}$ & Cassini experiment \cite{cassini}\tabularnewline
Perihelion advance & $(-1.8,\;1.4)\times10^{-3}$ & LAGEOS satellites \cite{Lucchesi:2010zzb}\tabularnewline
\; & $(-1.6,\;1.6)\times10^{-5}$ & MESSENGER mission \cite{Mercury}\tabularnewline
\; & $(-0.21,\;0.067)$ & Observation of S2 star at Galactic center \cite{s2}\tabularnewline
Geodetic precession & $(-2.6,\;2.6)\times10^{-3}$ & Gravity Probe B \cite{GPB}\tabularnewline
\; & $(-3.4,\;6.2)\times10^{-3}$ & Lunar laser ranging data \cite{lunar}\tabularnewline
Shadow (Rotating Solution) & $(-0.155,\;0.004)$ & Event Horizon Telescope Collaboration \cite{EHT1,EHT2,EHT3,EHT4}\tabularnewline
\end{tabular}
\end{ruledtabular}

\end{table*}

The stable circular orbit in the equatorial plane requires 
\begin{eqnarray}
E^{2}-V_{{\rm eff}}(R)=0\;\;\;{\rm and}\;\;\;\frac{dV_{{\rm eff}}}{dR}=0.
\end{eqnarray}
Solve these two equations, one has 
\begin{eqnarray}
E & = & \sqrt{\frac{g_{tt}^{2}(R)g'_{\phi\phi}(R)}{g'_{tt}(R)g_{\phi\phi}(R)-g_{tt}(R)g'_{\phi\phi}(R)}},\\
l & = & \sqrt{\frac{-g_{\phi\phi}^{2}(R)g'_{tt}(R)}{g'_{tt}(R)g_{\phi\phi}(R)-g_{tt}(R)g'_{\phi\phi}(R)}},\\
\Omega & = & \sqrt{-\frac{g'_{tt}(R)}{g'_{\phi\phi}(R)}}.\label{Omega}
\end{eqnarray}
Along this circular orbit, the parallel transport equation (\ref{spin_eq})
can be casted in the form of 
\begin{eqnarray}
 &  & \frac{ds^{t}}{d\lambda}+\frac{1}{2}\frac{g'_{tt}(R)}{g_{tt}(R)}u^{t}s^{R}=0,\label{eq1}\\
 &  & \frac{ds^{R}}{d\lambda}-\frac{1}{2}\frac{g'_{tt}(R)}{g_{RR}(R)}u^{t}s^{t}-\frac{1}{2}\frac{g'_{\phi\phi}(R)}{g_{RR}(R)}u^{\phi}s^{\phi}=0,\\
 &  & \frac{ds^{\theta}}{d\lambda}=0,\\
 &  & \frac{ds^{\phi}}{d\lambda}+\frac{1}{2}\frac{g'_{\phi\phi}}{g_{\phi\phi}}u^{\phi}s^{R}=0.
\end{eqnarray}
Differentiating (\ref{eq1}) with respect to the affine parameter
$\lambda$ and converting $\lambda\to t$ using the relation $dt=u^{t}d\lambda$
one arrives at a second-order ordinary differential equation of $s^{R}$,
\begin{equation}
\frac{d^{2}s^{R}}{dt^{2}}+\frac{1}{4}\left[\frac{g_{\phi\phi}'^{2}(R)}{g_{RR}(R)g_{\phi\phi}(R)}\Omega^{2}+\frac{g_{tt}'^{2}(R)}{g_{tt}(R)g_{\phi\phi}(R)}\right]s^{R}=0.
\end{equation}
This equation admits an exact solution 
\begin{eqnarray}
s^{R}(t)=s^{R}(0)\cos(\omega_{g}t),
\end{eqnarray}
where 
\begin{eqnarray}
\omega_{g}=\frac{1}{2}\sqrt{\frac{g_{\phi\phi}'^{2}(R)}{g_{RR}(R)g_{\phi\phi}(R)}\Omega^{2}+\frac{g_{tt}'^{2}(R)}{g_{tt}(R)g_{RR}(R)}},\label{omega}
\end{eqnarray}
represents the oscillating frequency pertaining to the spin four-vector
$s^{\mu}$. With the solution of $s^{R}$, the other three components
$s^{t}$, $s^{\theta}$, and $s^{\phi}$ can be immediately solved,
giving 
\begin{eqnarray}
s^{t}(t) & = & -\frac{1}{2}\frac{g'_{tt}(R)}{g_{tt}(R)}s^{R}(0)\sin(\omega_{g}t),\\
s^{\theta}(t) & = & 0,\\
s^{\phi}(t) & = & -\frac{1}{2}\frac{g'_{\phi\phi}(R)}{g_{\phi\phi}(R)}\Omega s^{R}(0)\sin(\omega_{g}t).
\end{eqnarray}
In obtaining these solutions, we have used initial conditions, $s^{t}(0)=s^{\theta}(0)=s^{\phi}(0)=0$,
which means spin vector $s^{\mu}$ was initially directed along the
radial direction.

Comparing Eqs.~(\ref{omega}) and (\ref{Omega}), it is obvious to
observe that the two frequencies, the oscillating frequency $\omega_{g}$
of rotation of the spin vector and the orbital frequency $\Omega$
of a massive spinning particle along the circular orbit, are different.
This difference leads to a precession of the spin vector. This is
the phenomenon called {\em geodetic precession}. For one complete
period of the circular orbit, the angle of the geodetic precession
can be expressed as 
\begin{eqnarray}
\Delta\Theta=2\pi\left(1-\frac{\omega_{g}}{\Omega}\right)\simeq\frac{3\pi G_{{\rm N}}M}{R}\left(1-\frac{2}{3}\frac{\eta+\tilde{k}}{1+\eta}\right).
\end{eqnarray}
When $\tilde{k}=0=\eta$, the above result reduces to the geodetic
precession of the Schwarzschild metric.

The geodetic precession can be tested and measured by using the gyroscopes
in near-earth artificial satellites. One such experiment is performed
by the Gravity probe B experiment, which was spaced at an altitude
of 642 km and had an orbital time period of 97.65 min. According to
GR, the geodetic effect induces a precession of the gyroscope spin
axis by 6,606.1 milliarcseconds (mas) per year. Gravity Probe B measures
this effect to be \cite{GPB} 
\begin{eqnarray}
\Delta\Theta=(6601.8\pm18.3){\rm mas}/{\rm year},
\end{eqnarray}
which leads to 
\begin{eqnarray}
-2.6\times10^{-3}<\frac{\eta+\tilde{k}}{2(1+\eta)}<2.6\times10^{-3}.
\end{eqnarray}
For the special Buchdahl-inspired metric, the above result gives 
\begin{eqnarray}
-2.6\times10^{-3}<\tilde{k}<2.6\times10^{-3}.
\end{eqnarray}

If one treats the Earth-Moon system as a gyroscope orbiting the Sun,
its geodetic precession due to the gravitational field of the Sun
has also been measured by using the Lunar laser ranging data. Recent
measurement of the geodetic precession yields a relative deviation
from GR as \cite{lunar} 
\begin{eqnarray}
\frac{\Delta\Theta-\Delta\Theta^{{\rm GR}}}{\Delta\Theta^{{\rm GR}}}=-0.0019\pm0.0064,\label{113}
\end{eqnarray}
which gives 
\begin{eqnarray}
-3.4\times10^{-3}<\frac{\eta+\tilde{k}}{2(1+\eta)}<6.2\times10^{-3}.
\end{eqnarray}
Again, this bound corresponds to 
\begin{eqnarray}
-3.4\times10^{-3}<\tilde{k}<6.2\times10^{-3},
\end{eqnarray}
for the special Buchdahl-inspired metric with $\eta=\tilde{k}$ and
$\zeta=\sqrt{1+3\tilde{k}^{2}}$.

\section{Conclusion\label{sec-con}}

In this paper, we study the observational constraints that can be
imposed on the asymptotically flat Buchdahl-inspired solution. For
this purpose, we calculate theoretically the effects of the parameter
$\tilde{k}$ on several solar system experiments and black hole observations.
Specifically, we calculate in detail the deflection angle of light
by the Sun, gravitational time delay, perihelion advance, and geodetic
procession for massless and massive objects in the Buchdahl-inspired
spacetime. With these theoretical predictions, we derive the constraints
on the parameter $\tilde{k}$ in the asymptotically flat Buchdahl-inspired
spacetime by confronting our theoretical calculations with observations.
Our results are summarized in Table~\ref{table}. In addition, we
have provided different comparisons of parameters from modified gravity
and the general relativity. For instance, Eq.~\eqref{47} provides
the comparison of deflection angle of light between two theories;
Eq.~\eqref{54} provides the comparison of gravitational time delay;
Eqs.~\eqref{75} and \eqref{82} provide similar comparison for perihelion
and periastron precession in both theories while Eq.~\eqref{113}
gives a comparison of geodetic precession as predicted by two theories.\vskip2pt

It is worth mentioning here that the measurement of the gravitational
time delay by the Cassini experiment provides the most sensitive tool
to constrain the parameter $\tilde{k}$ in the Solar System. Another
important constraint comes from observing the perihelion advance for
Mercury by the MESSENGER mission. As we mentioned, the measurement
of the anomalous perihelion advance for Mercury will be improved significantly
in the near future from the joint European-Japanese BepiColombo project,
which was launched in October 2018 \cite{Will:2018mcj,Bepi}. It
is expected that this mission will improve the accuracy of the perihelion
advance to be $10^{-4}$ as/century, which can be used to improve
the constraints on the parameters $\frac{\eta+\tilde{k}}{1+\eta}$
arising from the general Buchdahl-inspired metric or the parameter
$\tilde{k}$ in the special Buchdahl-inspired spacetime to $-10^{-6}\lesssim\frac{\eta+\tilde{k}}{1+\eta}\;{\rm or}\;\tilde{k}\lesssim10^{-6}$,
which is much more restricted than that obtained from the Cassini
experiment.\vskip2pt

In contrast to GR, pure ${\cal R}^{2}$ gravity does
not adhere to Birkhoff's theorem. As a higher-derivative characteristic,
the Buchdahl parameter $\tilde{k}$ of its vacuum solution exterior
to a star is system-dependent. Therefore, our empirical tests as presented
in this article were carried out under this premise. We have focused
on the exterior vacuum solution, deferring the theoretical determination
of $\tilde{k}$ for future exploration \cite{Nguyen-TOV-2024}.\vskip2pt

The determination of $\tilde{k}$ is, in principle,
contingent on the composition---specifically, the equation of state
and the distribution of matter within the host star. Typically, this
inquiry involves matching the interior and exterior solutions across
the star's surface. An alternative approach entails deriving a set
of Tolman-Oppenheimer-Volkoff (TOV) equations governing the pressure
and density of the star material. By numerically solving these equations
in conjunction with the metric components, the exterior vacuum configuration
of a star can be obtained based on a presumed equation of state and
conditions at the star's center \cite{Yagi-2021,Orellana-2013,Kase-2019}.
Progresses on this front have recently been made by one of us in reducing the TOV equations for $f({\cal R})$ gravity to a single integro-differential
equation. This simplication has enabled our investigation into the interior-exterior
matching for the Buchdahl-inspired solution, with detailed findings
to be reported separately \cite{Nguyen-TOV-2024}.\vskip2pt

While most test results presented in Table \ref{table}
align with GR, the cases for S2 star and M87{*}, which may qualify as
in a strong field regime, show large deviations for $\tilde{k}$ from
$0$, albeit with too large error bars for definitive conclusions; see Eqs. \eqref{S2-star} and \eqref{rotation3}. As investigated in \cite{2023-WH}, a $\tilde{k}$ value in the range
$(-1,0)$ has been associated with the potential formation of wormholes.
Theoretically, such spacetime configurations could support the possibility
of closed timelike curves, recently explored in \cite{Nguyen-CTC-2023}. Consequently,
future tests of the Buchdahl-inspired solution and pure $\mathcal{R}^{2}$
gravity in strong field regimes may be warranted.

\section*{Acknowledgments}

Tao Zhu is supported by the Zhejiang Provincial Natural Science Foundation
of China under Grants No. LR21A050001 and No. LY20A050002, the National
Natural Science Foundation of China under Grants No. 12275238 and
No. 11675143, the National Key Research and Development Program of
China under Grant No. 2020YFC2201503, and the Fundamental Research
Funds for the Provincial Universities of Zhejiang in China under Grant
No. RF-A2019015. Hoang Nguyen thanks Bertrand Chauvineau for stimulating discussions.

\end{document}